\documentclass[aip,preprint]{revtex4-2}
\usepackage{tikz}
\usepackage{pgfplots}
\usepackage{pgfplotstable}
\pgfplotsset{width=10cm,compat=1.9}
\usepackage[cp1251]{inputenc}
\usepackage[T2A]{fontenc}
\usepackage[english]{babel}
\usepackage{amssymb,latexsym,amsmath,amscd}
\usepackage{graphicx,color,framed}
\usepackage{bm}
\usepackage{appendix}
\usepackage{tabu}
\usepackage{tikz}
\usepackage{xcolor}
\usepackage{siunitx}
\usepackage{empheq}
\usepackage{float}
\usepackage{subcaption}
\usepackage{xr}
\usepackage{siunitx}
\usepackage{microtype}
\usepackage{amsfonts}
\usepackage{gensymb}
\usepackage{hyperref}
\usepackage[normalem]{ulem}
\graphicspath{{figures/}}

\makeatletter
\newcommand*{\addFileDependency}[1]{
  \typeout{(#1)}
  \@addtofilelist{#1}
  \IfFileExists{#1}{}{\typeout{No file #1.}}
}
\makeatother

\begin{document}

\author{\firstname{Victoria A.} \surname{Vasileva}}
\affiliation{Laboratory of Computational Physics, HSE University, Tallinskaya st. 34, 123458 Moscow, Russia}
\affiliation{School of Applied Mathematics, HSE University, Tallinskaya st. 34, 123458 Moscow, Russia}
\affiliation{Frumkin Institute of Physical Chemistry and Electrochemistry Russian Academy of Sciences, 31-4, Leninsky Prospect, 119071 Moscow, Russia}
\author{\firstname{Daria A.} \surname{Mazur}}
\affiliation{Laboratory of Computational Physics, HSE University, Tallinskaya st. 34, 123458 Moscow, Russia}
\author{\firstname{Petr E.} \surname{Brandyshev}}
\affiliation{Laboratory of Computational Physics, HSE University, Tallinskaya st. 34, 123458 Moscow, Russia}
\affiliation{School of Applied Mathematics, HSE University, Tallinskaya st. 34, 123458 Moscow, Russia}
\author{\firstname{Yury A.} \surname{Budkov}}
\email[]{ybudkov@hse.ru}
\affiliation{Laboratory of Computational Physics, HSE University, Tallinskaya st. 34, 123458 Moscow, Russia}
\affiliation{School of Applied Mathematics, HSE University, Tallinskaya st. 34, 123458 Moscow, Russia}
\affiliation{Laboratory of Multiscale Modeling of Molecular Systems, G.A. Krestov Institute of Solution Chemistry of the Russian Academy of Sciences, 153045, Akademicheskaya st. 1, Ivanovo, Russia}

\title{Osmolyte-Modulated Differential Capacitance and Disjoining Pressure for Nanoconfined Electrolytes: A Modified Poisson-Boltzmann Theory}

\begin{abstract}
This study employs modified Poisson-Boltzmann theory to systematically investigate the influence of zwitterionic osmolyte additives to an electrolyte solution on disjoining pressure and electric differential capacitance within charged slit-like nanopores with conductive walls. We demonstrate that increasing concentrations of zwitterionic osmolytes synergistically enhance both disjoining pressure and differential capacitance, highlighting their dual role in the potential improvement of supercapacitor performance. The insights gained underscore the unique capabilities of zwitterionic osmolytes as multifunctional additives for fine-tuning the properties of electric double layers, thereby bridging the gap between capacitive efficiency and microporous electrode longevity.
\end{abstract}

\maketitle

\section{Introduction}

Liquid-phase electrolytes confined within nanoscale porous electrodes are central to advancing electrochemical energy storage technologies, particularly supercapacitors~\cite{augustyn2014pseudocapacitive,wang2020electrode,simon2020perspectives,li2014koh,volfkovich2021electrochemical}. These systems rely on the electric double layer (EDL) formed at the interface between a conductive porous electrode and an electrolyte, where rapid ion adsorption enables high power density and cyclability. In supercapacitors, the EDL's structure within carbon micropores~\cite{li2014koh,koczwara2017situ,shao2020nanoporous} dictates critical performance metrics such as energy storage capacity and differential capacitance -- a key parameter defining the charge stored per unit voltage~\cite{kornyshev2007double,fedorov2014ionic}. The nanoconfinement of ions in these pores amplifies interfacial phenomena, where steric effects, ion correlations, and dielectric saturation diverge sharply from bulk behavior, directly influencing the EDL's thermodynamics and, consequently, the device's efficiency~\cite{kondrat2023theory}.

A paramount challenge in optimizing supercapacitors lies in quantifying the disjoining (or, more generally, solvation~\cite{gor2017adsorption,kolesnikov2021models}) pressure (stress) -- the mechanical force per unit area exerted by confined electrolytes on pore walls~\cite{derjaguin1987surface}. This solvation stress influences on the structural stability of porous electrodes during charge-discharge cycles \cite{gogotsi2011true,lukatskaya2013cation,shao2015graphene} and induces electrode deformation due to the electrosorption of ions \cite{augustyn2020deformation,koczwara2017situ,kolesnikov2022electrosorption,kolesnikov2025electrosorption}. Moreover, the disjoining pressure modulates ion packing density within pores~\cite{kondrat2023theory}, directly impacting the EDL's differential capacitance.  Moreover, disjoining (solvation) pressure is a crucial characteristic that describes the mechanical force generation by a carbon electromechanical nanoactuators when voltage is applied~\cite{baughman1999carbon,gor2017adsorption,gor2024drives}.

Accurate prediction of disjoining pressure remains a complex task due to competing nanoscale effects, including electrostatic correlations, solvation forces, and ion size limitations, which traditional mean-field models like Poisson-Boltzmann (PB) theory fail to capture. Standard PB frameworks, while foundational for modeling EDLs, neglect finite ion size~\cite{kornyshev2007double,borukhov1997steric,maggs2016general}, dielectric saturation~\cite{coalson1996statistical,abrashkin2007dipolar,gongadze2011langevin,iglivc2010excluded}, ion solvation effects~\cite{jiang2012solvent,ben2011dielectric,hatlo2012electric}, electrostatic correlations~\cite{netz2001electrostatistics,lue2006variational,wang2010fluctuation,huang2018confinement}, structural effects~\cite{blossey2017structural,de2020continuum,vasileva2023theory}, short-range specific interactions of species~\cite{goodwin2017mean}, electrode roughness effects~\cite{khlyupin2023molecular,nesterova2024role}, and internal electric structure of ions and solvent molecules~\cite{jiang2011density,jiang2012solvent,qing2021quantifying,buyukdagli2016beyond,slavchov2014quadrupole,slavchov2014quadrupole_2,budkov2020statistical,bossa2024incorporation}, and, finally, specific adsorption of ions~\cite{uematsu2018effects,podgornik2018general,wang2021demystifying,vinogradova2024surface} and image-charge effects~\cite{wang2013effects,zhou2024image} rendering them rather inaccurate for nanoconfined systems. Modified Poisson-Boltzmann (MPB) equations address these gaps by incorporating non-mean-field corrections, asymmetric ion partitioning, and surface interactions (see, \cite{blossey2023poisson,budkov2024statistical} and references therein). For supercapacitors, these models based on the MPB theory may be essential for designing porous electrodes with optimized pore geometries and surface chemistry to maximize the differential capacitance~\cite{kondrat2023theory}. Recently, significant efforts have been devoted to advanced molecular simulations of ion adsorption on real carbon electrodes immersed in aqueous electrolytes~\cite{misra2019theory}, as well as to the calculation of differential capacitance of the EDL~\cite{seebeck2022elucidating,wang2025constant} and the disjoining pressure in slit-like pores~\cite{schlaich2018simulations,kolesnikov2025electrosorption}.

An important yet underexplored issue is the role of zwitterionic additives -- molecules bearing spatially separated positive and negative charges -- in modulating disjoining pressure and EDL thermodynamics and dynamics. In energy storage contexts, zwitterions could enhance ion accessibility within micropores by preventing aggregation and optimizing charge distribution, thereby increasing differential capacitance. Their dipolar nature introduces localized charge regulation and screening effects~\cite{govrin2018zwitterionic,govrin2019common}, potentially reshaping confinement thermodynamics. While zwitterions are established as biocompatible stabilizers in biological systems (e.g., osmolyte-mediated membrane stabilization~\cite{imamura2018synthesis,warepam2024osmolyte}), their impact on electrochemical systems remains speculative. Specifically, their ability to alter disjoining pressure and ion structuring in nanopores could unlock strategies to mitigate electrode degradation and enhance differential capacitance.

In a recent paper~\cite{bruch2024model}, the authors examined the theory of EDL that forms at the metal-solution of zwitterionic polymers interface. These unique polymers feature oppositely charged sites linked by flexible, neutral linear chains. By employing self-consistent field theory, the authors revealed that, from the perspective of supercapacitor applications, these systems concurrently exhibit the properties of electrolytes and dielectric materials. This dual nature leads to the formation of EDL and facilitates the passage of an electric field through the system. Furthermore, the authors speculated on the potential of hybrid supercapacitors that incorporate salt solutions along with zwitterionic polymer additives. Such devices could function effectively at low electrode potentials, where ion movement significantly contributes to capacitance formation, and at high potentials, where the movement of zwitterions enhances capacitance. Additionally, it is important to consider low-molecular-weight zwitterionic co-solvents, particularly betaine oligopeptides. This is crucial because these relatively small molecules can more easily penetrate the nanoscale pores of supercapacitor electrodes than polymeric ones with high conformational entropy~\cite{budkov2023macroscopic,kondrat2023theory}. In addition, due to their significant dipole moments, betaine molecules could sharply increase the differential capacitance when they are present in electrolyte solutions, due to the enhancement of the local dielectric permittivity~\cite{frydel2011polarizable}. Consequently, investigation of electrolyte solutions that contain admixtures of these zwitterionic molecules, especially in the confined geometries of electrified nanopores, is an essential task for advancing the fields of modern chemical physics and electrochemistry.

This study uses the MPB formalism to investigate the disjoining pressure and the differential capacitance behavior of the EDL in charged slit-like nanopores filled with electrolyte solutions containing additives of a zwitterionic co-solvent. We show that increasing the concentration of zwitterions enhances the electrostatic disjoining pressure. Moreover, we demonstrate that the addition of zwitterions to an electrolyte solution results in a simultaneous increase in differential capacitance. These findings have potential implications for the design of supercapacitors.

\section{Theoretical background}
\begin{figure}[H]
\centering\includegraphics[height=10.0 cm]{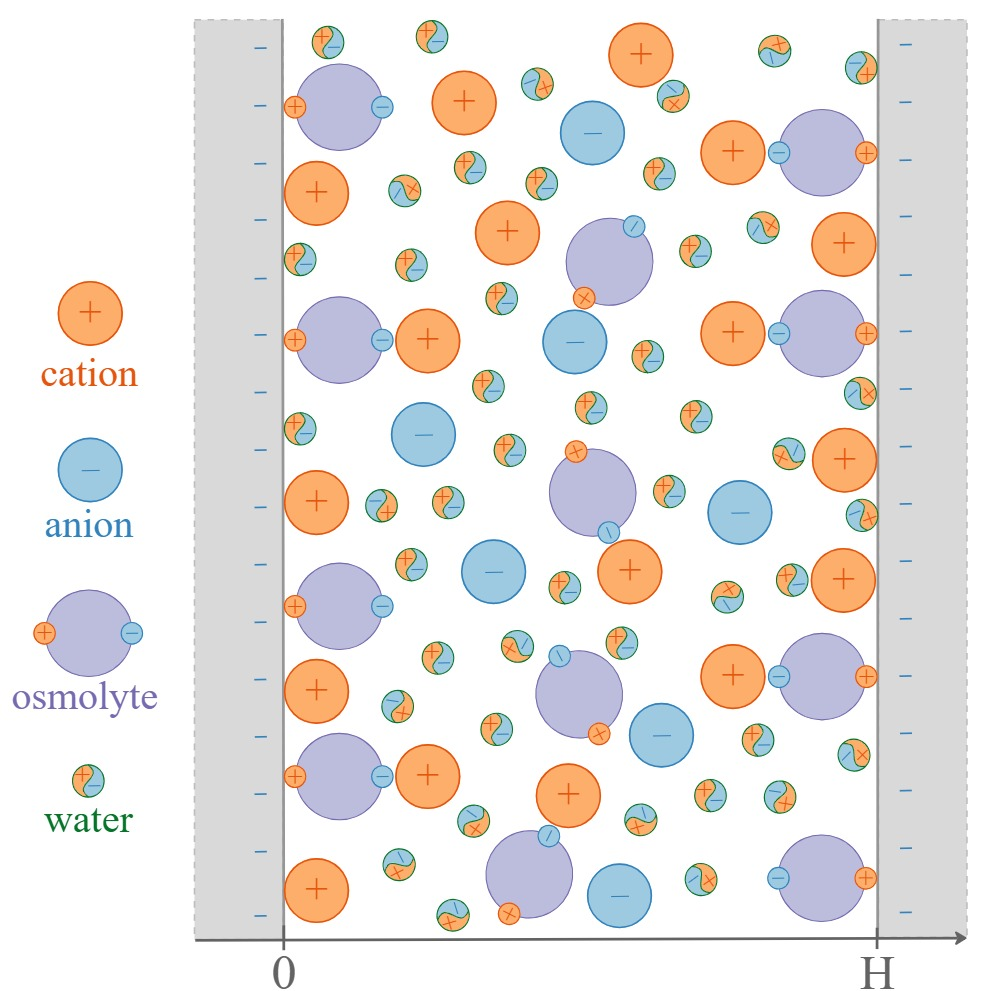}
\caption{Illustration of the current model -- a confined aqueous electrolyte solution with an admixture of zwitterionic additives.}
\label{Picture}
\end{figure}

Let us consider a 1:1 aqueous electrolyte solution with an admixture of the zwitterionic molecules confined within an infinite slit-like pore, with electrified conductive walls of a porous electrode, separated by a distance $H$ (Fig. \ref{Picture}). We assume that the confined solution is in equilibrium with the bulk solution, so that the grand thermodynamic potential per unit wall's surface area can be written in the form~\cite{budkov2024statistical}
\begin{multline}
\label{omega}
\Omega=\int\limits_0^{H} d z\bigg{(}-\frac{\varepsilon_0E^2}{2}+\rho \psi\bigg{)}+ k_B T \int\limits_0^{H} d z \left(c_{\text{z}} \ln \frac{\beta p_{\text{z}} E}{\sinh \beta p_\text{z} E}+c_{\text{w}} \ln \frac{\beta p_{\text{w}} E}{\sinh \beta p_{\text{w}} E}\right) \\
+\int\limits_0^{H} d z\left(f\left(c_\text{w},c_\text{z}, c_{+}, c_{-}\right)-\sum_\alpha c_\alpha (\mu_{\alpha}-w_\alpha)\right),
\end{multline}
where the first term in the right hand side of eq. (\ref{omega}) is the electrostatic energy of ions (without water molecules and zwitterions) in the mean-field approximation with $E(z)=-\psi'(z)$ being the local electric field; $c_{\pm}$ are the local ionic concentrations; $\psi(z)$ is the local electrostatic potential and $\rho(z)=q(c_{+}(z)-c_{-}(z))$ is the local charge density of ions with charges $\pm q$ ($q=e$ is the elementary charge); $\varepsilon_0$ is the permittivity of vacuum. The second term is the free energy of two-component dipole gas in "external" self-consistent field $\bold{E}$ \cite{budkov2018theory}; $c_{\text{w}}(z)$ is the local concentration of water molecules with the dipole moment $p_\text{w}$ and $c_\text{z}(z)$ is the local concentration of the zwitterions with dipole moment $p_\text{z}$. The third term is the grand thermodynamic potential of solution within the reference fluid~\cite{budkov2024statistical} model taking into account the interaction of the species with the walls; $w_{\alpha}(z)=u_{\alpha}(z)+u_{\alpha}(H-z)$ is the total potential energy of interaction between species $\alpha$ and the walls; $u_{\alpha}(z)$ in turn is the single-wall potential which will be specified below; $k_B$ is the Boltzmann constant, $T$ is the temperature, $\beta = (k_B T)^{-1}$. In this paper we use the hard-spheres mixture as the reference fluid system with the following free energy density in the Percus-Yevick approximation~\cite{roth2010fundamental}
\begin{equation}
f=f_{\text{id}}+f_{\text{hs}},
\end{equation}
where the first term describes the ideal gas contribution to the total free energy density
\begin{equation}
f_{\text{id}}=k_{B}T\sum\limits_{\alpha}c_{\alpha}\left(\ln(c_{\alpha}\Lambda_{\alpha}^3/\xi_\alpha)-1\right),
\end{equation}
with $\Lambda_{\alpha}$ being the thermal wavelengths of the species and $\xi_\alpha$ -- their internal partition functions, which will drop out from the final result; the second term describes the contribution of steric interactions of the species within the model of hard sphere mixture:
\begin{equation}
\label{PY}
\frac{f_{\text{hs}}}{k_B T} = -n_0 \ln(1-n_3) + \frac{n_1 n_2}{1 - n_3} + \frac{n_2^3}{24\pi(1-n_3)^2}
\end{equation}
The auxiliary variables are: $n_0 =\sum_{\alpha} c_{\alpha}$, $n_1 = 1/2 \sum_{\alpha} c_{\alpha}d_{\alpha}$, $n_2 = \pi\sum_{\alpha} c_{\alpha}d_\alpha^2$, $n_3 = \pi/6  \sum_{\alpha}c_{\alpha}d_{\alpha}^3$, where $d_{\alpha}$ are diameters of the hard spheres.

The self-consistent field equations can be obtained from the grand thermodynamic potential variation, i.e. from the following Euler-Lagrange equations
\begin{equation}
\frac{\delta \Omega}{\delta \psi(z)}=0,\quad \frac{\delta \Omega}{\delta c_\alpha(z)}=0,\quad \alpha=\pm/\text{z}/\text{w},
\end{equation}
which yield the Poisson equation
\begin{equation}
\label{MPB}
\frac{d}{dz}\left(\epsilon(z) \frac{d}{dz} \psi(z)\right)=-q\left(c_{+}(z)-c_{-}(z)\right),
\end{equation}
with the local permittivity of the solution~\cite{coalson1996statistical,abrashkin2007dipolar,budkov2022modified}
\begin{equation}
\label{permit}
\epsilon(z)=\varepsilon_0 \left(1+\frac{p_{\text{w}}^2}{ \varepsilon_0 k_B T} \frac{L(\beta p_{\text{w}} E(z))}{\beta p_{\text{w}} E(z)} c_{\text{w}}(z)+\frac{p_{\text{z}}^2}{ \varepsilon_0 k_B T} \frac{L(\beta p_{\text{z}} E(z))}{\beta p_{\text{z}} E(z)} c_{\text{z}}(z)\right),
\end{equation}
where $L(x)=\coth(x)-{1}/{x}$ is the Langevin function, and the chemical equilibrium equations for the local concentrations of species
\begin{equation}
\label{eq_for_conc1}
\bar{\mu}_{ \alpha}+w_{\alpha} + q_{\alpha} \psi = \mu_{ \alpha},~\alpha = \pm
\end{equation}
 \begin{equation}
\label{eq_for_conc2}
\bar{\mu}_{\alpha} +w_\alpha+k_B T \ln \frac{\beta p_{\alpha} E}{\sinh \beta p_{\alpha} E} = \mu_{\alpha},~\alpha =\text{w}/\text{z},
\end{equation}
where
\begin{equation}
\label{mu}
\bar \mu_\alpha =  \dfrac{\partial f}{\partial c_\alpha},~\alpha=\pm,\text{w}/\text{z}
\end{equation}
is the intrinsic chemical potential of the species $\alpha$. The bulk chemical potentials, $\mu_{\alpha}$, are determined by eqs. (\ref{mu}), with $c_+ = c_- = c$, $c_\text{w} = c_{\text{w}, b}$, and $c_\text{z} = c_{\text{z}, b}$, where $c$ is the ion concentration in the bulk, $c_{\text{w}, b}$ is the bulk water concentration, and $c_{\text{z}, b}$ is the bulk zwitterion concentration.

To obtain the local concentrations of ions and zwitterions, we need to solve a system of coupled equations (\ref{MPB}) and (\ref{eq_for_conc1}), (\ref{eq_for_conc2}), with the boundary conditions $\psi(0) = \psi(H) = \psi_0$, where $\psi_0$ is the surface potential.

To calculate the differential capacitance, we need to know the wall's surface charge density, $\sigma=-\epsilon(0)\psi'(0)$, where $\epsilon(0)$ is the permittivity on the wall surface. The differential capacitance is $C=d\sigma /d\psi_0$.

The disjoining pressure can be calculated by \cite{budkov2022modified,budkov2024thermomechanical}
\begin{equation}
\label{Pi2}
\Pi=P_{m}-P_{b}-2\sum\limits_{\alpha}\int\limits_{H/2}^{H}dz c_{\alpha}(z)u_{\alpha}^{\prime}(z),
\end{equation}
where $P_{m}=P\left(H/2\right)$ is the normal pressure at $z=H/2$ generated by the species. The local osmotic pressure is $P=\sum_{\alpha} c_{\alpha}\bar{\mu}_{\alpha}-f$ and $P_{b}$ is the bulk pressure, i.e. pressure, $P$, at $c_{\alpha}=c_{\alpha,b}$. For a relatively large pore, where the single-wall potential range is much smaller than the pore width, as is the case in this study, we can ignore the integral on the right-hand side of eq. (\ref{Pi2}) and use the Langmuir equation~\cite{derjaguin1987surface}, $\Pi = P_m - P_b$, to calculate the disjoining pressure.

\section{Numerical results and discussions}
In this section, we present the results of our calculations regarding the electrical differential capacitance and disjoining pressure for a confined aqueous electrolyte containing zwitterionic additives. These results are derived from the self-consistent field equations (\ref{MPB}), (\ref{eq_for_conc1}), and (\ref{eq_for_conc2}). To obtain $\epsilon=78\varepsilon_0$ for pure water ($c=c_{\text{z},b}=0$) at $T=298~$K, following references~\cite{abrashkin2007dipolar,mazur2024understanding}, we assume the effective dipole moment $p_{\text{w}}=4.8~$D at $c_{\text{w},b}=55~$M. The effective water diameter is $d_{\text{w}}= 0.282~$ nm. We will focus on zwitterionic glycine in both peptide and dipeptide forms. For zwitterions, we assume that the hard sphere diameter can be calculated as $d_\text{z} = p_\text{z}/q$, that is, we assume that charged sites are located on the sphere diametrically opposed to each other (see Fig.~\ref{Picture}). In the following, we will consider glycine in its peptide form with a dipole moment of $p_z=17$ D and in its dipeptide form with $p_z = 27.6$ D~\cite{pethig1984dielectric}. We assume that the hard sphere diameters to be the twice as high the hydration radii of ions, i.e. $d_{+}=2\times 0.331~$nm\cite{nightingale1959phenomenological} and $d_{-}=2\times 0.383~$nm~\cite{mazur2024understanding}. These values approximately correspond to the aqueous solution of KPF$_6$~\cite{mazur2024understanding}. Our choice of an aqueous electrolyte solution is due to the fact that the model of a mixture of hard spheres for it gives a qualitatively correct behavior of the differential capacitance ~\cite{huang2023density,mazur2024understanding} and does not lead to overestimated capacitance values, as in the case of the lattice gas model~\cite{budkov2021theory}. In principle, this model could be improved by considering effects such as the dielectric decrement of ions\cite{ben2011dielectric} and specific ion-electrode, water-electrode, and ion-water interactions. This would allow us to align theoretical capacitance profiles with experimental data, as demonstrated in our recent work~\cite{mazur2024understanding}. However, we have chosen not to do so in this paper in order to keep the focus on the main message, limiting ourselves to demonstrating the physical effects without being tied to specific ions or electrodes.

We will also consider the case of specific adsorption of species modelling the single-wall potentials as the Gaussian-well potentials~\cite{vasileva2023theory}
\begin{equation}
\label{eq:gaussian}
u_{\alpha}(z)=\epsilon_{\alpha}\exp\left[-\frac{z^2}{2\sigma_{\alpha}^2}\right],
\end{equation}
where $\epsilon_{\alpha}<0$ is the potential depth and $\sigma_{\alpha}$ is the potential range. Note that it is not necessary to attach a deep meaning to the functional form of the selected adsorption potentials. To simulate a potential well, we can also choose other potentials, such as a square-well potential~\cite{uematsu2018effects,budkov2018theory}. The Gaussian-well potential is convenient for calculations due to its smooth behavior and the ability to easily adjust the depth and range of interaction.

Note that in this study we consider sufficiently wide pores ($H>1~\text{nm}$), as in this case, we may not need to take into account the contribution of solvation forces to the disjoining pressure associated with the strong heterogeneity of water~\cite{hedley2023dramatic,kolesnikov2023water}. Additionally, we consider rather dilute solutions of electrolytes and zwitterions, so that non-local effects that are important for ionic liquids and concentrated solutions, can be ignored. Nevertheless, we will consider concentrations of zwitterions that are much higher than the concentrations of the electrolyte in the bulk phase. This is because the effect of adding electrically neutral, dipolar zwitterions is expected to have a much weaker influence on the differential capacitance and disjoining pressure than the addition of ions with electric charges.

\subsection{Differential capacitance}
\subsubsection{No specific adsorption.}
\begin{figure}[H]
\centering
\begin{subfigure}[t]{0.48\linewidth}
    \includegraphics[height=6.cm]{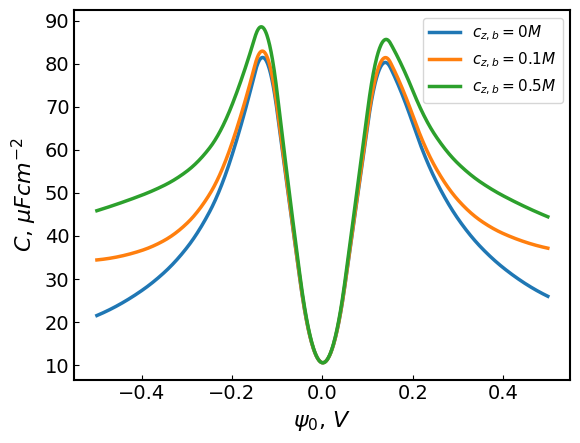}
    \caption{peptide, ($p_{\text{z}}=17~\text{D}$)}
   \label{fig1a}
 \end{subfigure}
\hfill
\begin{subfigure}[t]{0.48\linewidth}
    \includegraphics[height=6.cm]{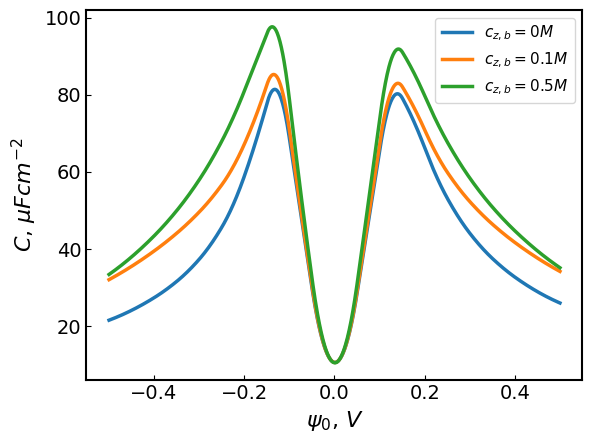}
    \caption{dipeptide, ($p_{\text{z}}=27.6~\text{D}$)}
    \label{fig1b}
\end{subfigure}

\caption{Differential capacitance profiles are plotted for different glycine bulk concentrations in peptide (a)  and dipeptide (b) forms and a fixed bulk electrolyte concentration $c=0.01~\text{M}$. The data are shown for, $T=298~\text{K}$, $H=3~\text{nm}$, $\epsilon_{\text{w}}=\epsilon_{\text{z}}=\epsilon_{\pm}=0$.} 
\label{Capacitance}
\end{figure}

\begin{figure}[H]
\centering
\begin{subfigure}[t]{0.48\linewidth}
    \includegraphics[height=6.cm]{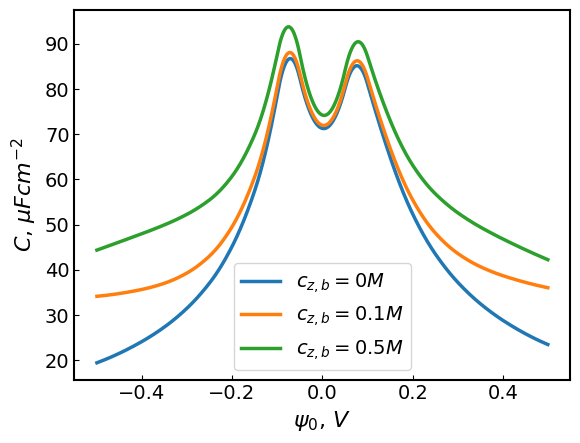}
    \caption{peptide, ($p_{\text{z}}=17~\text{D}$)}
   \label{fig2a}
 \end{subfigure}
\hfill
\begin{subfigure}[t]{0.48\linewidth}
    \includegraphics[height=6.cm]{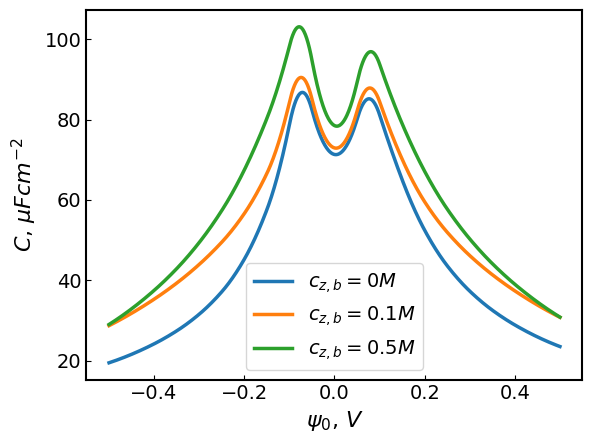}
    \caption{dipeptide, ($p_{\text{z}}=27.6~\text{D}$)}
    \label{fig2b}
\end{subfigure}

\caption{Differential capacitance profiles are plotted for different zwitterionic osmolyte (glycine) bulk concentrations in peptide (a)  and dipeptide (b) forms and a fixed bulk electrolyte concentration $c=0.1~\text{M}$. The data are shown for, $T=298~\text{K}$, $H=3~\text{nm}$, $\epsilon_{\text{w}}=\epsilon_{\text{z}}=\epsilon_{\pm}=0$.} 
\label{Capacitance_2}
\end{figure}

\begin{figure}[H]
\centering
\begin{subfigure}[t]{0.48\linewidth}
    \includegraphics[height=6.cm]{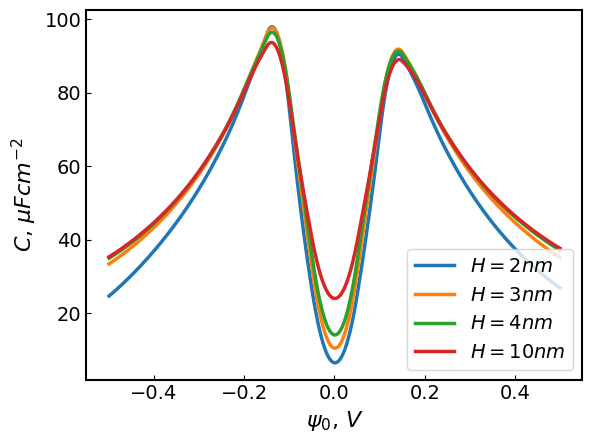}
    \caption{dilute solution, ($c=0.01~\text{M}$)}
   \label{pore_effect_1}
 \end{subfigure}
\hfill
\begin{subfigure}[t]{0.48\linewidth}
    \includegraphics[height=6.cm]{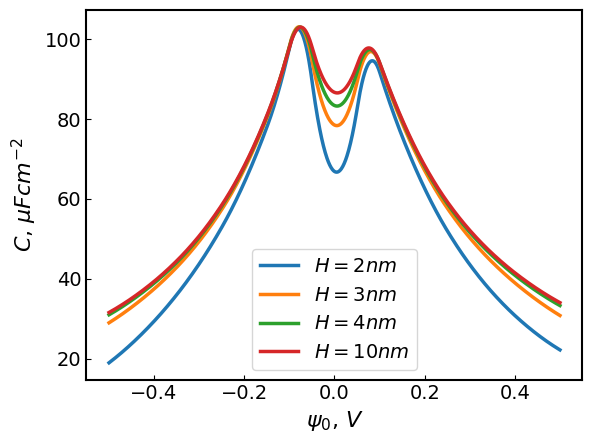}
    \caption{moderate solution, ($c=0.1~\text{M}$)}
    \label{pore_effect_2}
\end{subfigure}

\caption{Differential capacitance profiles are plotted for different pore widths in dilute solution (a)  and moderate solution (b). The data are shown for $T=298~\text{K}$, $\epsilon_{\text{w}}=\epsilon_{\text{z}}=\epsilon_{\pm}=0$.} 
\label{Pore_effect}
\end{figure}

Fig. \ref{Capacitance} shows the differential capacitance profile of the confined aqueous electrolyte in a pore with a thickness of $H=3$ nm for a rather dilute electrolyte concentration $c_s=0.01$ M, at different glycine concentrations in both peptide and dipeptide forms. As can be seen, an increase in the concentration of peptide and dipeptide glycines leads to a slight increase in differential capacitance at relatively low surface potentials ($\psi_0\sim \pm 0.1$ V) and a more significant increase at higher concentrations ($\psi_0\sim \pm 0.2-0.5$ V). Note that the effect persists for moderately concentrated electrolyte solutions ($c=0.1$ M) (see Fig. \ref{Capacitance_2}). Note that we only show data for one pore size, $H=3$ nm, as the differential capacitance changes only slightly for pores of size $H > 2.5$ nm at rather large surface potential ($\psi_0\sim \pm 0.2-0.5$ V). Furthermore, changes in pore size in this range do not violate the effect of differential capacitance enhancement discussed (see Fig. \ref{Pore_effect}). As illustrated in Fig. \ref{Pore_effect}, at relatively low potentials, pore size significantly influences differential capacitance; specifically, smaller pores result in a deeper minimum of differential capacitance. This behavior is attributed to the overlapping of electric EDLs in narrower pores, which complicates ion penetration. Consequently, this leads to a reduction in differential capacitance.

\begin{figure}[H]
\centering\includegraphics[height=7.5 cm]{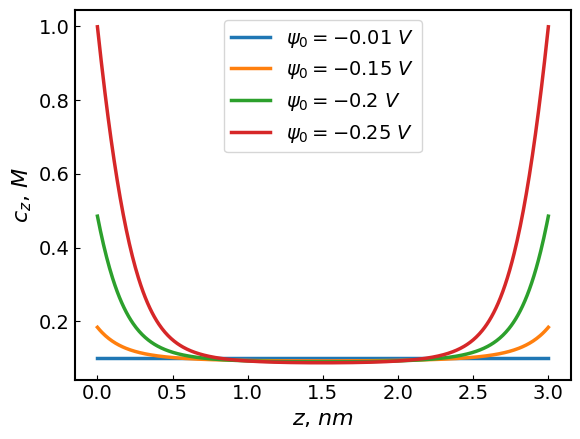}
\caption{Dipeptide glycine concentration profiles plotted for different electrode potentials. The data are shown for $c=0.01~\text{M}$, $c_{\text{z},b} = 0.1$ M, $T=298~\text{K}$, $p_{\text{z}}=27.6~\text{D}$, $H=3~\text{nm}$, $\epsilon_{\text{w}}=\epsilon_{\text{z}}=\epsilon_{\pm}=0$.}
\label{Concentrations}
\end{figure}
 
\begin{figure}[H]
\centering\includegraphics[height=7 cm]{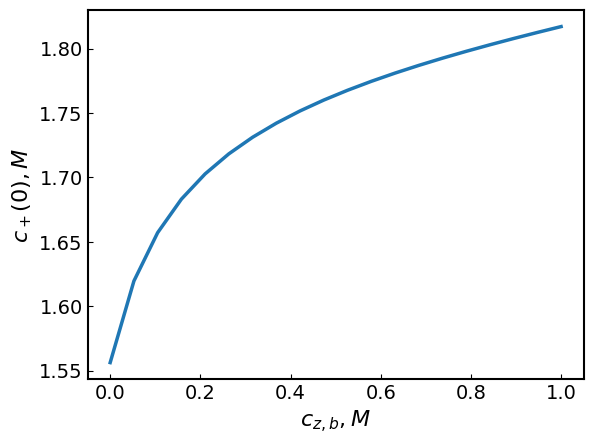}
\caption{Cation local concentration at the boundary ($z = 0$) plotted for different dipeptide glycine concentrations. The data are shown for $c=0.01~\text{M}$, $T=298~\text{K}$, $p_{\text{z}}=27.6~\text{D}$, $\psi_0 = -0.2~\text{V}$, $H=4~\text{nm}$, $\epsilon_{\text{w}}=\epsilon_{\text{z}}=\epsilon_{\pm}=0$.}
\label{Local concentration}
\end{figure}

The significant increase in differential capacitance observed in the high-potential region can be attributed to two primary factors. Near the electrode interface, the charge from the electrode is effectively screened by mobile ions, resulting in a highly inhomogeneous electric field that significantly influences on the polar glycine molecules. This electric field induces a dielectrophoretic force that draws the glycine molecules toward the electrode surface~\cite{budkov2018theory}. Consequently, there is an increased concentration of glycine near the electrode (as illustrated in Fig. \ref{Concentrations}), which subsequently enhances the local permittivity of the solution. This is the first direct factor contributing to the enhancement of differential capacitance. The second, indirect factor is that the increase in permittivity leads to a greater accumulation of ionic charge at elevated surface potentials within the electric double layer (EDL), driven by the electrostatic attraction between ionic charges and the dipole moments of the glycine molecules (see Fig. \ref{Picture}). Indeed, as demonstrated in Fig. \ref{Local concentration}, the concentration of cations at the electrode monotonically increases with the bulk dipeptide glycine concentration. This effect is relatively negligible at low surface potentials, but becomes pronounced when the potential exceeds a certain threshold (approximately ( $|\psi_0| \sim 0.15$ V). 

\begin{figure}[H]
\centering
\begin{subfigure}[t]{0.48\linewidth}
    \includegraphics[height=6.cm]{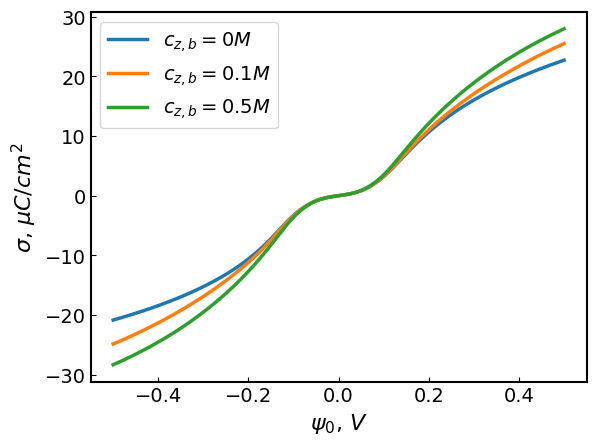}
    \caption{dilute solution, ($c=0.01~\text{M}$)}
   \label{sigma_1}
 \end{subfigure}
\hfill
\begin{subfigure}[t]{0.48\linewidth}
    \includegraphics[height=6.cm]{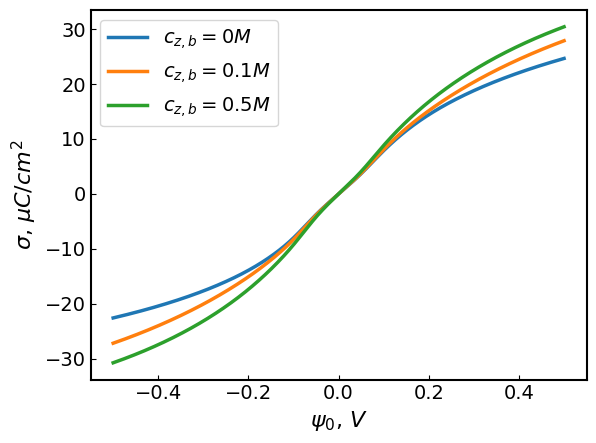}
    \caption{moderate solution, ($c=0.1~\text{M}$)}
    \label{sigma_2}
\end{subfigure}

\caption{Surface charge density profiles are plotted for different bulk zwitterion concentrations. At a rather large surface potential, an increase in the bulk concentration of zwitterions results in an increase in the surface charge density on the electrode. The data are shown for $T=298~\text{K}$, $c=0.01~$M, $p_{\text{z}}=27.6~\text{D}$, $\epsilon_{\text{w}}=\epsilon_{\text{z}}=\epsilon_{\pm}=0$.} 
\label{sigma}
\end{figure}

It is noteworthy that this effect is also reflected in the behavior of the electrode's surface charge density, as shown in Fig. \ref{sigma}. Specifically, an increase in the bulk zwitterion concentration enhances the surface charge density of the electrode at relatively high absolute values of surface potential.

\begin{figure}[H]
\centering
\begin{subfigure}[t]{0.48\linewidth}
    \includegraphics[height=6.cm]{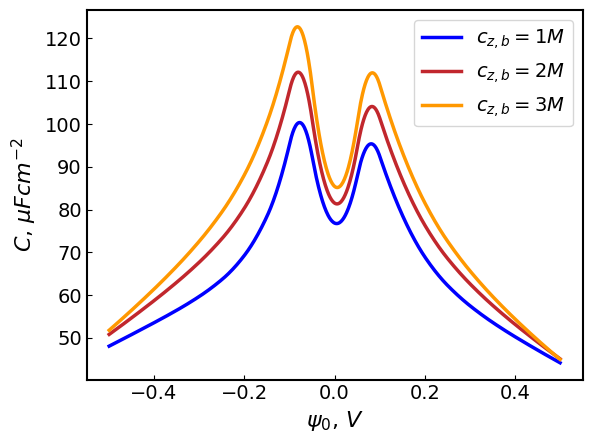}
    \caption{peptide glycine ($p_{\text{z}}=17~\text{D}$)}
   \label{fig2a}
 \end{subfigure}
\hfill
\begin{subfigure}[t]{0.48\linewidth}
    \includegraphics[height=6.cm]{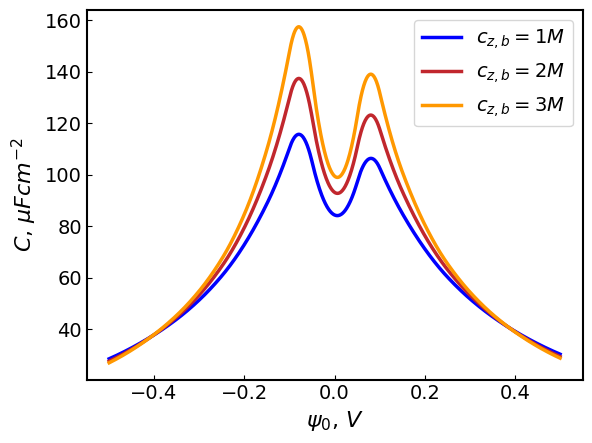}
    \caption{dipeptide glycine ($p_{\text{z}}=27.6~\text{D}$)}
    \label{fig2b}
\end{subfigure}

\caption{Differential capacitance profiles are plotted for different zwitterionic osmolyte (glycine) bulk concentrations in peptide (a) and dipeptide (b) forms and a fixed bulk electrolyte concentration $c=0.1~\text{M}$. Present theory predicts a substantial effect on the differential capacitance for both peptide and dipeptide forms of glycine. The data are shown for $T=298~\text{K}$, $H=3~\text{nm}$, $\epsilon_{\text{w}}=\epsilon_{\text{z}}=\epsilon_{\pm}=0$.} 
\label{Capacitance_predictions}
\end{figure}

It is instructive to consider the accessible zwitterion concentrations in the experiment that produce the most pronounced osmolytic effect on differential capacitance, as discussed above. Recent experiments~\cite{govrin2018zwitterionic,govrin2019common} have shown that realistic osmolyte solubilities of $1-3~\text{M}$ in aqueous electrolytes can be readily achieved. Fig.\ref{Capacitance_predictions} illustrates a significant increase in differential capacitance for a moderate electrolyte solution ($c=0.1$ M) as the bulk glycine concentration rises from $1~\text{M}$ to $3~\text{M}$. As observed, present theory predicts a substantial effect for both peptide and dipeptide forms of glycine. This prediction should be verified experimentally in the future.

\subsubsection{Effects of specific adsorption of different species.}
\begin{figure}[H]
\centering\includegraphics[height=7.5 cm]{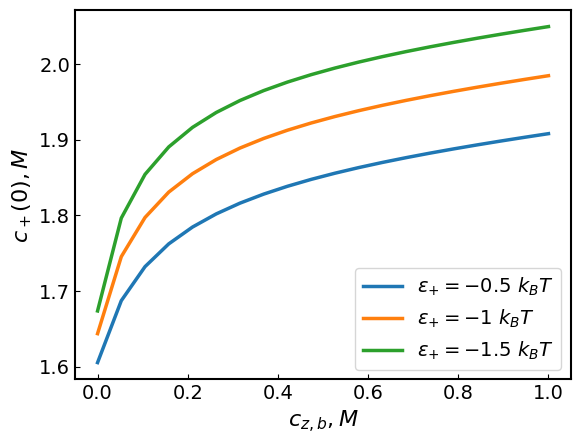}
\caption{Cation concentration at the boundary ($z = 0$) as a function of different dipeptide glycine concentrations. Specific adsorption of cations is accounted with parameters in the Gaussian-well adsorption potential range $\sigma_+ = d_+$ and different $\epsilon_+$. The data are shown for $c=0.01~\text{M}$, $T=298~\text{K}$, $p_{\text{z}}=27.6~\text{D}$, $\psi_0 = -0.2~\text{V}$, $H=4~\text{nm}$, $\epsilon_{-}=\epsilon_{\text{z}}=\epsilon_{\text{w}}=0$.}
\label{Local concentration_2}
\end{figure}

It is important to note that a similar local cation concentration enhancement occurs when considering the specific adsorption of cations onto the electrode surface (refer to Fig. \ref{Local concentration_2}). Our calculations suggest that the charge enhancement in the EDL, induced by the additive of zwitterionic osmolytes into the electrolyte solution, is a generic physical phenomenon that can be qualitatively interpreted within the framework of the MPB theory.

\begin{figure}[H]
\centering\includegraphics[height=7.5 cm]{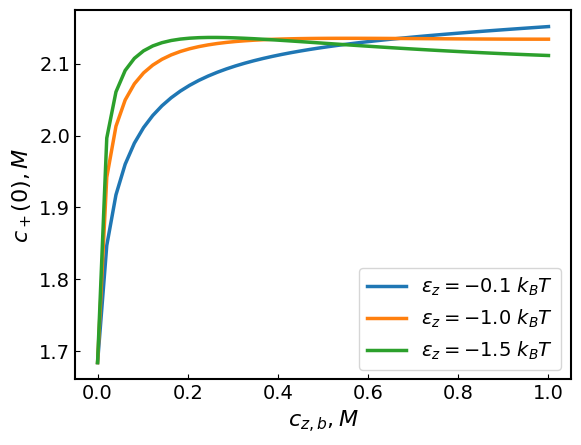}
\caption{Cation concentration at the boundary ($z = 0$) as a function of dipeptide glycine concentrations. Specific adsorption of zwitterion is accounted with parameters in the Gaussian-well adsorption potential range $\sigma_{\text{z}} = d_{\text{z}}= 0.55~\text{nm}$ and different $\epsilon_{\text{z}}$. The data are shown for $c=0.01~\text{M}$, $T=298~\text{K}$, $p_{\text{z}}=27.6~\text{D}$, $\psi_0 = -0.3~\text{V}$, $H=4~\text{nm}$, $\epsilon_{-}=\epsilon_{\text{z}}=\epsilon_{\text{w}}=0$.}
\label{Local concentration_3}
\end{figure}

It is also interesting to study how the local concentration of ions on the electrode changes with an increase in the concentration of zwitterions in the solution. This is because the specific adsorption of these molecules onto the electrode can lead to a complex behavior of their local concentration. As can be seen in Fig. \ref{Local concentration_3}, with sufficiently strong adsorption, the local concentration does not always increase monotonously with the bulk zwitterion concentration. Instead, it first increases, reaches a maximum, and then decreases relatively slowly. This is due to the fact that when there is a large number of zwitterionic molecules near the electrode, they expel ions from the surface. 

\begin{figure}[H]
\centering\includegraphics[height=7.5 cm]{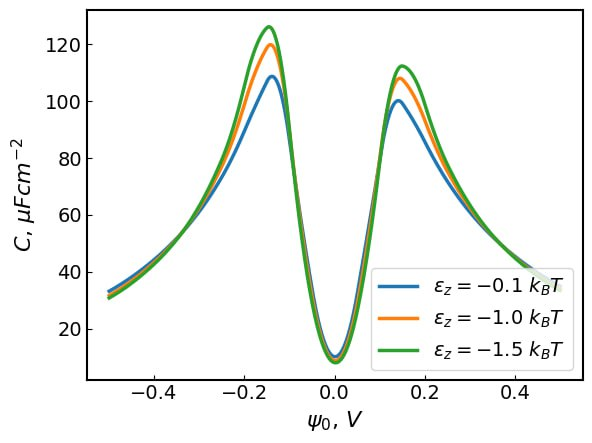}
\caption{Differential capacitance profiles, plotted for different zwitterion adsorption parameters, $\epsilon_{\text{z}}$. The data are shown for $T=298~\text{K}$, $c=0.01~$M, $c_{\text{z},b}=0.9~\text{M}$, $p_{\text{z}}=27.6~$D, $H=4~\text{nm}$, $\sigma_{\text{z}}=d_{\text{z}}=0.55~$nm,  $\epsilon_{\pm}=\epsilon_{w}=0$.}
\label{Diff_capacitance}
\end{figure}

Interestingly, despite the reduction in local concentration of ions on the electrode surface caused by their replacement with the zwitterions, the differential capacitance actually grows with increase of the zwitterion adsorption absolute value (see Fig. \ref{Diff_capacitance}). Such net effect emerges because the local permittivity enhancement due to highly polar zwitterions influx outweighs the ionic charge depletion from the double layer.

\subsection{Disjoining pressure}

\begin{figure}[H]
\centering
\begin{subfigure}[t]{0.48\linewidth}
    \includegraphics[height=6.cm]{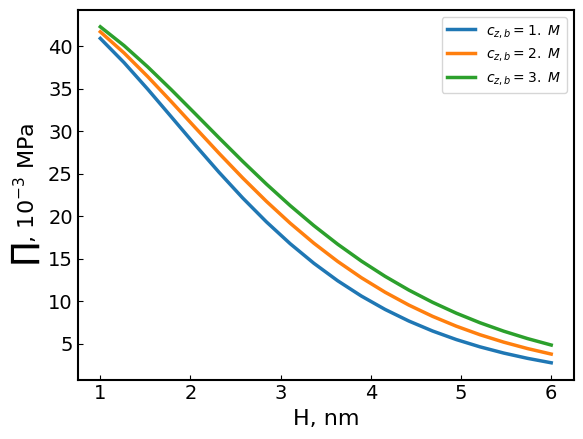}
    \caption{$\psi_0 = -0.015~\text{V}$}
   \label{sigma_1}
 \end{subfigure}
\hfill
\begin{subfigure}[t]{0.48\linewidth}
    \includegraphics[height=6.cm]{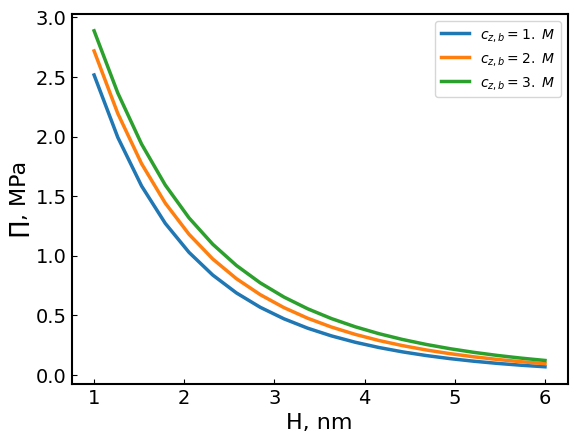}
    \caption{$\psi_0 = -0.1~\text{V}$}
    \label{sigma_2}
\end{subfigure}
\caption{Disjoining pressure as a function of the pore size $H$, plotted for different concentrations of glycine in peptide form at rather small (a) and rather large surface potential (b). The data are shown for $c=0.05~\text{M}$, $T=298~\text{K}$, $p_{\text{z}}=17~\text{D}$, $\epsilon_{\text{w}}=\epsilon_{\text{z}}=\epsilon_{\pm}=0$.} 
\label{DP}
\end{figure}

Recent experimental studies by Govrin et al.~\cite{govrin2018zwitterionic} revealed that addition of the osmolytes to an electrolyte solution diminishes the electrostatic screening of charged surfaces by the mobile ions present in the solution, thereby amplifying the repulsive forces between them. In the context of slit-like pores with charged walls, this phenomenon is expected to result in an enhancement of the electrostatic component of the disjoining pressure as the concentration of zwitterionic molecules in the bulk solution increases. The MPB theory predicts this behavior, as illustrated in Figures \ref{DP}(a,b). It suggests the rise of the disjoining pressure with increase of glycine concentration, both in its peptide and dipeptide form. Notably, this effect is observed at two distinct ranges of surface potentials: at low surface potentials ($\psi_0=-0.015~\text{V}$ in Fig.\ref{DP}a), where the glycine concentration within the pore resembles that of the bulk solution, and at higher potentials ($\psi_0=-0.1~\text{V}$ in Fig.\ref{DP}b), where the glycine concentration near the wall exceeds that of the bulk. In both scenarios, the impact is more pronounced for glycine dipeptides compared to peptides. This enhanced effect can be attributed to a substantial increase in the local permittivity of the solution, which reduces walls' charge screening by the mobile ions and intensifies electrostatic repulsion of the charged walls. In the case of relatively large surface potentials, where an increase in dipeptide glycine concentration leads to a significant increase in differential capacitance, as well as weakening of electrostatic screening in pores, the disjoining pressure increases due to the fact that the local glycine concentration within these pores is higher than in the surrounding solution. This leads to a higher osmotic pressure at the center of the pore compared to that in the bulk solution, not only due to the ionic contribution, as in the case of small potentials, but also due to the presence of excess osmolyte molecules. 

\begin{figure}[H]
\centering\includegraphics[height=7.5 cm]{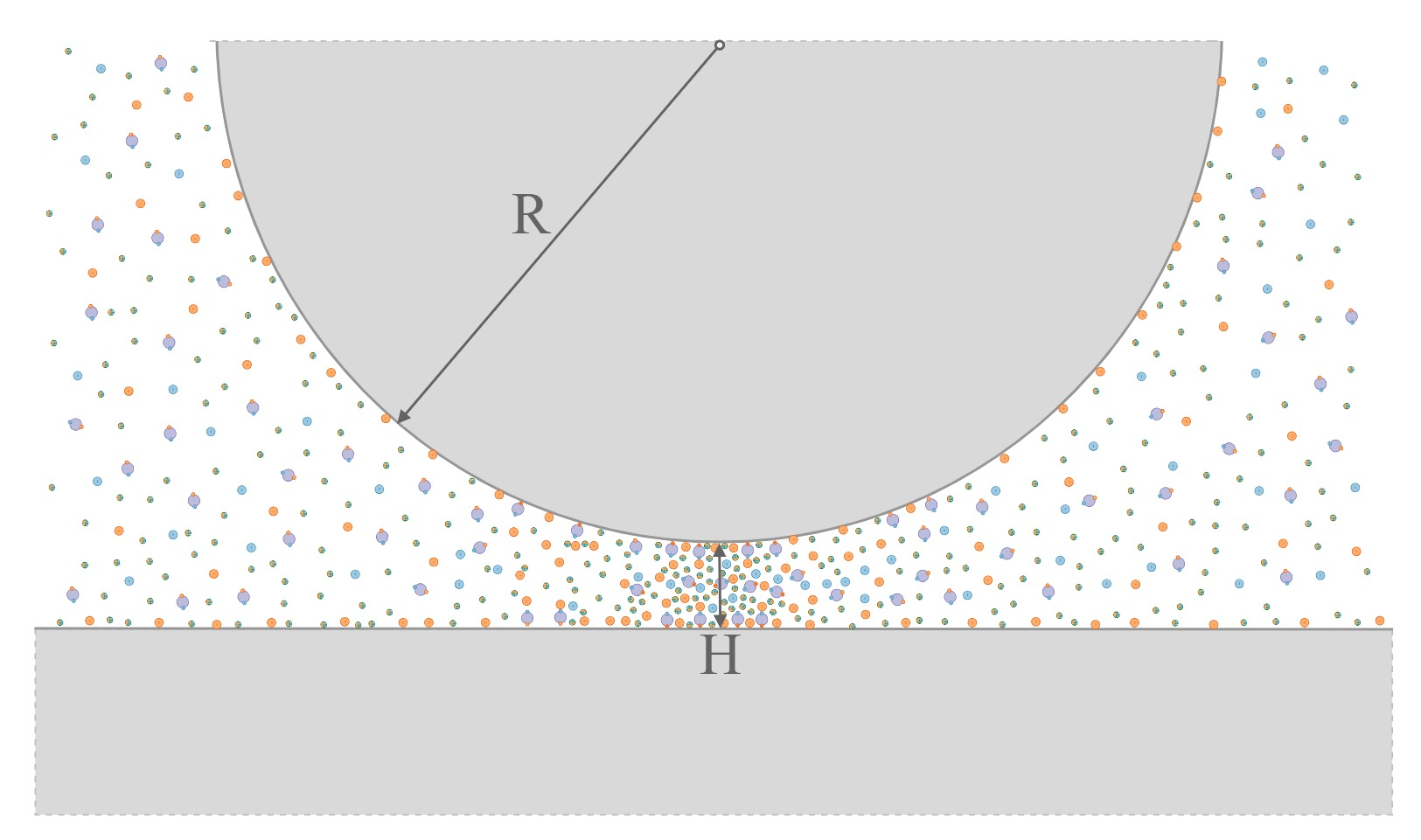}
\caption{A scheme of a hypothetical atomic force microscope showing a cantilever and a metal substrate of identical metal under the same electrostatic potential immersed in an aqueous electrolyte solution containing zwitterions.}
\label{AFM}
\end{figure}

\begin{figure}[H]
\centering\includegraphics[height=7.5 cm]{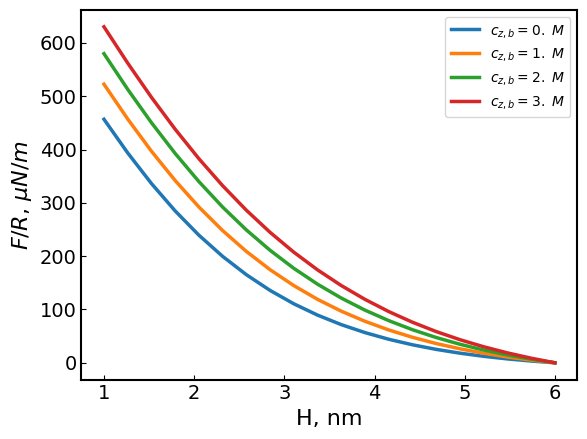}
\caption{The force of the electrostatic repulsion between the conductive cantilever and the substrate of the same metal, divided by catilever radius, when placed in an aqueous electrolyte solution containing glycine impurities, as a function of their mutual separation, in the Derjaguin approximation.  The data are shown for $c=0.05~\text{M}$, $\psi_0=-0.015~\text{V}$, $T=298~\text{K}$, $p_{\text{z}}=17~\text{D}$, $\epsilon_{\pm}=\epsilon_{\text{z}}=\epsilon_{\text{w}}=0$.}
\label{force}
\end{figure}

Finally, it would be reasonable to make a qualitative comparison with the mentioned experiment~\cite{govrin2018zwitterionic}. Strictly speaking, within present consideration we cannot directly compare our results to the experiment where the charge on silica surfaces was regulated due to cation desorption caused by zwitterion additives. To directly compare, it would be necessary to formulate a charge regulation model~\cite{podgornik2018general}, which is outside the scope of this paper. Instead, we consider a hypothetical scenario of a spherical metal cantilever from an atomic force microscope interacting with a flat substrate of identical metal under the same surface potential in an aqueous electrolyte solution containing glycine in peptide form (see Fig.\ref{AFM}). We assume that the electrostatic potential is sufficiently small to approximately correspond to the values of the surface charge density on the substrate accessible in Govrin et al.'s experiment. The force can be estimated using the Derjaguin's approximation~\cite{derjaguin1987surface}
\begin{equation}
F(H)=2\pi R\int\limits_{H}^{\infty} dx \Pi(x),
\end{equation}
where $R$ represents the cantilever radius and $\Pi(x)$ represents the disjoining pressure embedded in the slit-like pore with a thickness of $x$. As can be seen in Fig.\ref{force}, an increase in the bulk zwitterion concentration from 0 to $3~$M results in a significant increase in the electrostatic repulsive force between the cantilever and the substrate, which is similar to what was observed in the experiments by Govrin et al.~\cite{govrin2018zwitterionic}

\section{Discussion}

Now, we would like to speculate about possible implications of the predicted effects for practical applications. \textbf{First}, it is particularly noteworthy that incorporating zwitterionic osmolytes into an electrolyte solution can lead to a substantial enhancement in differential capacitance, even at low electrolyte concentrations. This characteristic presents a significant advantage in electrochemical applications, as it effectively reduces unwanted electrochemical activity of ions at the electrode surface, without loss of differential capacitance. By partially substituting these ions with zwitterionic osmolytes the overall electrochemical performance of the system is improved. Their neutral charge prevents participation in Faradaic reactions, thereby reducing parasitic electrochemical activity that typically plagues traditional electrolytes. This suppression of ionic activity may enhance electrode stability by mitigating parasitic side reactions that degrade its long-term performance. \textbf{Second}, during the operation of the supercapacitors, the micropores found within the electrodes are susceptible to collapse due to the electrode ageing ~\cite{huang2017experimental,erradi2021performance}. Such structural degradation can severely compromise the performance and efficiency of the supercapacitor. In contrast, zwitterionic osmolytes induce osmolytic effects that strengthen "intra-pore" repulsive forces, thereby reducing the likelihood of collapse and preserving the structural integrity of the micropores. Maintaining this integrity may be important, as it ensures that the porous electrodes retain a high surface area available for charge storage. A higher surface area correlates directly to improved capacitance and energy storage capabilities. Consequently, the introduction of zwitterionic osmolytes may not only stabilize the pore structure but also contribute to maintaining consistent capacitance levels over extended periods, ultimately enhancing the reliability and performance of supercapacitors in various applications.

Finally, it is instructive to discuss the possible limitations of present theoretical model. It is important to note that this model neglects the electrostatic correlations between ions and polar molecules of glycine and water. When implicitly accounting for the solvent using a constant permittivity of the solvent medium, it is known that correlation effects, including those arising from image forces, can be addressed within the framework of variational field theory through the electrostatic Green's function~\cite{wang2010fluctuation,wang2013effects,lue2006variational,zhou2024image}. However, when explicitly considering the polar solvent or co-solvent molecules, solving the equations of variational field theory becomes significantly more challenging. Tackling this complex numerical problem will allow for a more precise refinement of the effects predicted in this study, which is certainly a topic for further research.

\section{Conclusion}
In conclusion, we utilized modified Poisson-Boltzmann theory to systematically explore the role of zwitterionic osmolyte additives in modulating the disjoining pressure and differential capacitance of electric double layers within charged slit-like nanopores with conductive walls. Our findings revealed that increase of zwitterion concentration induces a pronounced synergistic enhancement of both disjoining pressure and differential capacitance, a dual effect potentially useful for optimizing supercapacitor performance. We believe that our findings reveal the unique potential of zwitterionic osmolytes as multifunctional additives capable of tailoring electric double layer properties, bridging the gap between capacitive efficiency and electrode durability. By elucidating the molecular mechanisms underlying this synergy, our work provides a foundational framework for designing advanced supercapacitors that harmonize high energy density with long-term operational stability.

\textbf{Data availability statement.} {\sl The data that supports the findings of this study are available within the article.}

\section*{Acknowledgments}
The authors thank the Russian Science Foundation (Grant No. 24-11-00096) for financial support of parts of this paper devoted to the theoretical background and disjoining pressure calculations. The part on the osmolyte's effect on the differential capacitance was partly supported by the Ministry of Science and Higher Education of the Russian Federation.

\selectlanguage{english}
\bibliography{name}

\end{document}